\documentclass[aps,revtex,prb,twocolumn,amsmath,showpacs,floatfix,reprint]{revtex4-2}
\usepackage{amsmath, nccmath}
\usepackage{amssymb}
\usepackage{bm}
\usepackage{braket}
\usepackage{natbib}
\usepackage{leftidx}
\usepackage{graphicx}    
\usepackage{verbatim}   
\usepackage{color}      
\usepackage{subfigure}  
\usepackage{hyperref}   
\usepackage{dcolumn}    
\usepackage{textcomp}
\usepackage{float}
\usepackage{threeparttable}
\usepackage{titlecaps}
\usepackage{multirow}
\usepackage{lipsum}
\hypersetup{
	colorlinks=true,
	citecolor=blue,
	filecolor=black,
	linkcolor=blue,
	urlcolor=cyan
}
\usepackage{mathastext}
\usepackage{mathptmx}
\usepackage{enumitem}
\DeclareGraphicsExtensions{.png,.pdf,.tif}
\usepackage{pict2e}
\usepackage[dvipsnames]{xcolor}
\usepackage[normalem]{ulem}

\begin{document}

\title{Spin-reorientation Driven Temperature Dependent Intrinsic Anomalous Hall Conductivity in Fe$_3$Ge, a Ferromagnetic Topological Metal}

\author{Susanta Ghosh}
\affiliation{Condensed Matter Physics and Material Sciences Department, S. N. Bose National Centre for Basic Sciences, Kolkata, West Bengal-700106, India}

\author{Tushar Kanti Bhowmik}
\affiliation{Condensed Matter Physics and Material Sciences Department, S. N. Bose National Centre for Basic Sciences, Kolkata, West Bengal-700106, India}

\author{Achintya Low}
\affiliation{Condensed Matter Physics and Material Sciences Department, S. N. Bose National Centre for Basic Sciences, Kolkata, West Bengal-700106, India}


\author{Setti Thirupathaiah}
\email{setti@bose.res.in}
\affiliation{Condensed Matter Physics and Material Sciences Department, S. N. Bose National Centre for Basic Sciences, Kolkata, West Bengal-700106, India}

\date{\today}

\begin{abstract}
We investigate the temperature dependence of the intrinsic anomalous Hall conductivity in Fe$_3$Ge, which is a ferromagnetic topological metal. We observe a significant anisotropy in the anomalous Hall conductivity between in-plane and out-of-plane directions. We further identify that the total Hall conductivity is contributed extrinsically due to the skew-scattering mechanism and intrinsically due to nonzero Berry curvature in the momentum space. Most importantly,  we demonstrate the temperature dependence of the intrinsic Hall contribution,  a rare phenomenon to visualize experimentally,  due to tuning the easy-magnetic axis from the out-of-plane to the in-plane with decreasing temperature. We also show that the extrinsic Hall conductivity decreases with temperature as  $\sigma_{xy}^{ext}(T)=\frac{\sigma_{xy0}^{ext}}{(aT+1)^2}$  due to electron-phonon scattering.

\end{abstract}

\maketitle

\section{Introduction}

The Hall effect, a cornerstone of solid-state physics, arises from Lorentz forces acting on charge carriers as they traverse in a conductor under perpendicular applied magnetic field~\cite{Kittel1955, Ashcroft1976}. This fundamental phenomenon has found widespread utility in various technological applications, from precise magnetic field sensing to the characterization of materials' electronic properties~\cite{Chien2013}. The conventional Hall effect usually occurs in non-magnetic metals~\cite{Ashcroft1976, Nagaosa2010}. In contrast, the magnetic materials exhibit an anomalous Hall effect (AHE), distinguished by a pronounced increase in Hall resistivity proportionate to the material's magnetization~\cite{Nagaosa2010}. The AHE was initially attributed to two extrinsic scattering processes, namely the skew-scattering and the side-jump mechanisms~\cite{smit1955, berger1970}. In 1954, Karplus and Luttinger (KL) proposed a different theory to understand the AHE by introducing a band structure-originated contribution to the AHE, an intrinsic contribution~\cite{Karplus1954}. It was later understood that the intrinsic KL mechanism is directly related to the Berry phase effects on the occupied electronic Bloch states~\cite{Jungwirth2002, Nagaosa2010}. Usually, the Berry phase in the momentum space is considered temperature-independent as the electronic band structure does not change much with temperature unless the system possesses temperature-dependent electronic, structural, or magnetic phase transitions. Therefore, not much information is available on the temperature-dependence of the intrinsic Hall contribution, except for one recent paper discussing the possibility of such effect~\cite{Ye2012}.

\begin{figure}[ht]
	\includegraphics[width=0.90\linewidth]{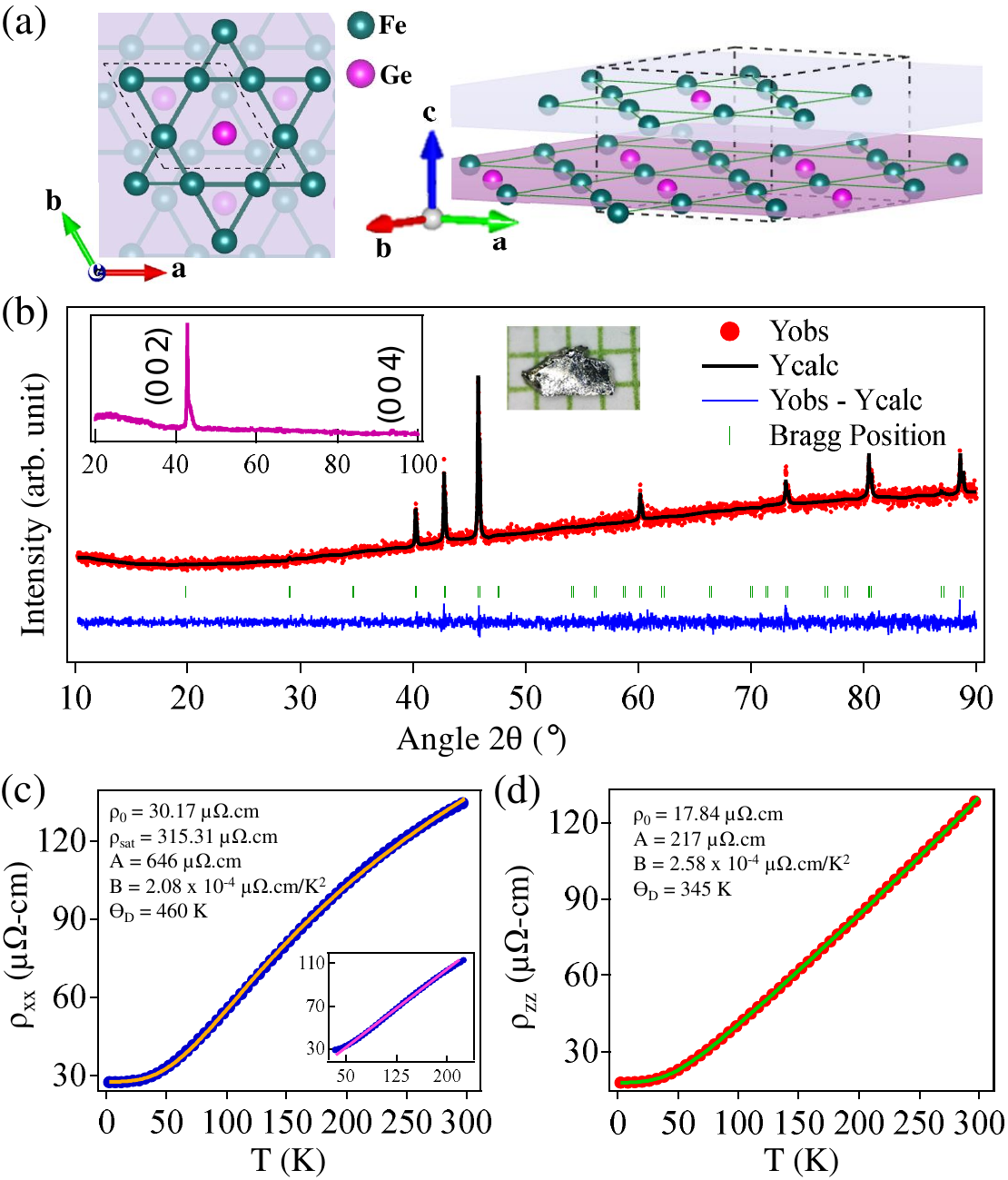}
	\caption{(a) Schematic crystal structure of the hexagonal unit cell and kagome lattice of $Fe_3Ge$. (b) Powder X-ray diffraction (XRD) pattern of the crushed single crystals of Fe$_3$Ge overlapped with the Rietveld refinement. Insets in (b) display a photographic image of the Fe$_3$Ge single crystal and XRD pattern taken on the Fe$_3$Ge single crystal. (c) and (d) depict the temperature dependent in-plane ($\rho_{xx}$) and out-of-plane ($\rho_{zz}$) electrical resistivity, respectively. The solid curves in (c) and (d) are fits using the Bloch-Wilson equation (see the text for more details). Inset in (c) shows the linear resistivity within the temperature range of 50-200 K.}
	\label{fig1}
\end{figure}

In this paper, the kagome lattice, consisting of two-dimensional corner-sharing triangles, enables distinctive electronic behaviors such as van Hove singularities, Dirac-like dispersion, and flat bands~\cite{Kang2019, Yin2022, Sau2024}. Its unique structural geometry combined with topology and electronic correlations~\cite{Yin2022} provide a rich platform for exploring emergent quantum phenomena. The Hall effect in kagome systems is currently a hotbed of research activities. These materials exhibit large AHE~\cite{Morali2019, Nakatsuji2015} and topological Hall effect (THE)~\cite{Wang2020}, driving the exploration for novel electronic properties and functionalities in these systems. Among them, the Co-based ferromagnetic kagome material Co$_3$Sn$_2$S$_2$ stands out for its giant AHE~\cite{Morali2019, Wang2018}. Interestingly, the Mn-based kagome compounds such as Mn$_3$X (X = Sn, Ge), despite being antiferromagnetic, demonstrate a large AHE, attributed to the intrinsic nonzero Berry curvature~\cite{Nakatsuji2015, Nayak2016, Kuebler2018}. The ferromagnetic kagome system \( Fe_3Sn_2 \) exhibits both AHE and THE~\cite{Hou2017, Ye2019, Wang2020}, while the ferromagnetic compound \( Fe_3Sn \) displays a substantial AHE, with contributions originated from both intrinsic and extrinsic sources~\cite{Low2023}.

On the other hand, the composition \( \mathrm{Fe}_3\mathrm{Ge} \), similar to \( \mathrm{Fe}_3\mathrm{Sn} \), has been reported to exhibit room-temperature ferromagnetism and large anomalous Hall effect (AHE)~\cite{Drijver1976, Li2023}. Most importantly, unlike Fe$_3$Sn, Fe$_3$Ge shows the tuning of easy-magnetization axis from the out-of-plane to the in-plane below the spin-reorientation transition at around 380 K~\cite{Drijver1976, Albertini2004, Zhang2024}. Such a spin-reorientation transition significantly changes the electronic band structure~\cite{Lou2024}, providing an ideal platform to study the temperature effect on the intrinsic Hall conductivity. Therefore, we synthesized the single crystals of \( \mathrm{Fe}_3\mathrm{Ge} \) and thoroughly investigated its directional dependent magnetic and Hall effect properties. Magnetic measurements indicate that below \( T_{SR} \), the Fe magnetic moments gradually cant from the \( z \)-direction towards the \( xy \)-plane as the temperature decreases. We observe that the Hall conductivity has both intrinsic and extrinsic contributions. Importantly,  we demonstrate the temperature dependence of intrinsic Hall contribution due to tuning of the easy-magnetic axis from the out-of-plane to the in-plane with decreasing temperature. We further discuss the effect of electron-phonon scattering on the extrinsic Hall conductivity originating from the skew-scattering mechanism.

\begin{figure*}[ht]
\includegraphics[width=\linewidth, clip=true]{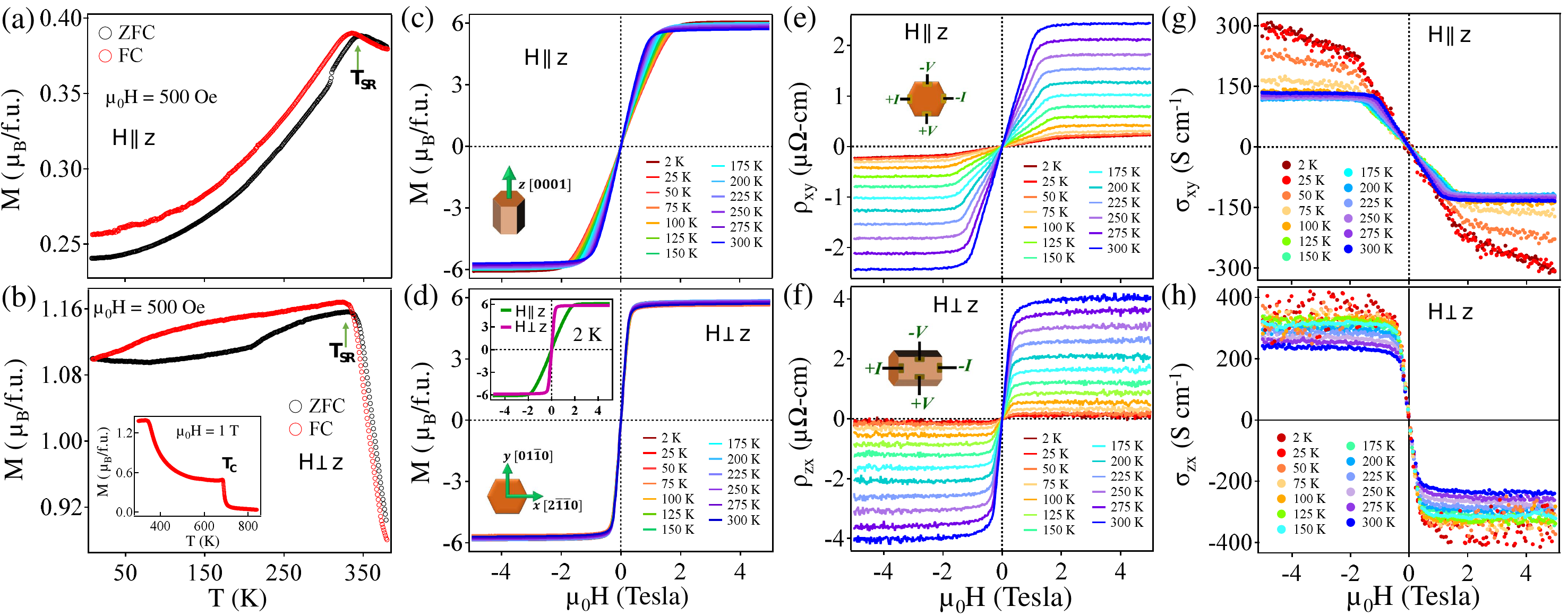}
\caption{Temperature-dependent magnetization [$M(T)$] of Fe$_3$Ge plotted for (a) $H\parallel z$ and (b) $H\perp z$ under an applied magnetic field of $H=$ 500 Oe. Inset in (b) shows the $M(T)$ data at high temperatures. Magnetization isotherms [$M(H)$] plotted for (c) $H\parallel z$ and (d) $H\perp z$ at various sample temperatures. In-plane ($\rho_{xy}$) and out-of-plane ($\rho_{zx}$) Hall resistivity plotted as a function of the field for (e) $H\parallel z$ and (f) $H\perp z$  at various sample temperatures. In-plane ($\sigma_{xy}$) and out-of-plane ($\sigma_{zx}$) Hall conductivity plotted as a function of the field for (g) $H\parallel z$ and (h) $H\perp z$  at various sample temperatures.}
\label{fig2}
\end{figure*}

\section{METHODOLOGY}
\subsection{Experimental details}

Single crystals of Fe$_{3}$Ge were prepared by melt growth technique. In this method, high-purity iron (Alfa Aesar 99.999$\%$) and germanium (Alfa Aesar 99.999$\%$) powders were thoroughly ground in an argon-filled glove box to form a homogeneous mixture. The mixture was then transferred into an alumina crucible, which was subsequently sealed within a preheated quartz ampoule under a vacuum. The sealed ampoule was heated to \(1120^\circ\)C in a high-temperature box furnace and kept at that temperature for a day for a thorough chemical reaction. Subsequently, the temperature was slowly reduced to \(800^\circ\)C at a controlled cooling rate of \(2^\circ\)/hour. After prolonged annealing for the next five days at \(800^\circ\)C, the sample was rapidly quenched in ice water to prevent the formation of low-temperature phases.

Phase purity and structural characterization were done using powder X-ray diffraction (XRD) with Cu$K_{\alpha}$ radiation (9 kW) on a Rigaku X-ray diffractometer. Chemical composition was identified using energy-dispersive X-ray spectroscopy (EDS) measurements. The EDS data reveal that the exact chemical composition of the as-prepared samples is Fe$_{3.04}$Ge$_{0.96}$,  which is very close to the nominal composition of Fe$_{3}$Ge. Further, the EDS elemental mapping verifies the homogeneity and compositional uniformity of the single crystal studied single crystals. Scanning electron microscope (SEM) image reveals no evidence of grain boundaries within the crystal, confirming the sample's single-crystalline nature [see Fig.~S1 in Ref.~\cite{Supple}].

Electrical transport and magnetic properties were measured using a 9-Tesla physical property measurement system (PPMS, Dynacool, Quantum Design) over a broad temperature range between 2 and 380 K. High-temperature magnetic behavior was explored with a VSM-oven attached to the PPMS. Hall effect and transport measurements employed the standard four-probe method, with copper leads meticulously attached to the sample using H20E/1Oz silver epoxy for reliable contact. For convenience, the cartesian coordinates (\(x, y, z\)) are used instead of crystallographic axes (\(a, b, c\)) wherever required in this manuscript. We can identify the \(c\)-axis (\(z\)) and \(ab\)-plane (\(xy\)) using XRD measurements, but we were unable to differentiate between the \(a\)- and \(b\)-axes. Therefore, to streamline the manuscript, the longer side of the crystal's \(xy\)-plane (inset of Fig.~\ref{fig1}(b)) is considered as the \(x\)-axis, and the shorter side as the \(y\)-axis.

\subsection{First-principles Calculations}
The electronic ground states of $Fe_3Ge$ were calculated using density functional theory (DFT) within the Perdew-Burke-Ernzerhof (PBE) generalized gradient approximation (GGA), implemented in the Quantum Espresso (QE) simulation package \cite{Giannozzi2009}. The crystal structure was optimized with ultrasoft pseudo-potentials~\cite{Vanderbilt1990}, using the force and energy convergence criteria of $10^{-4}$ $Ry/\AA$ and $10^{-5}$ Ry, respectively. An energy cutoff of 110 Ry and a charge density cutoff of 1320 Ry were applied for the plane-wave basis set. A $20\times20\times20$ k-point mesh was employed for Brillouin zone sampling. The Methfessel-Paxton (mp) smearing technique with a smearing parameter of $\sigma=0.007$ Ry was utilized for the charge density evaluation. The spin-orbit coupling effect was incorporated using the fully relativistic pseudo-potentials. To explore the anomalous Hall conductivity (AHC), a tight-binding model was constructed using maximally localized Wannier functions (MLWFs) via the Wannier90 code~\cite{Mostofi2014}. The Berry curvature along high-symmetry directions was computed using the Kubo formolism~\cite{Thouless1982} as also implemented in Wannier90 code. The intrinsic AHC along the [001] and [010] directions were then obtained by integrating the $z$- and $y$-components of the Berry curvature over the entire Brillouin zone using the WannierTools code~\cite{Wu2017}.

\section{Results and Discussions}

The left panel in Fig.~\ref{fig1}(a) depicts the schematic view of the hexagonal crystal structure of Fe$_3$Ge and the right panel in Fig.~\ref{fig1}(a) shows crystal structure projected onto $ab$-plane from which we can identify the kagome lattice plane formed by the Fe atoms with hexagons and triangles and Ge sitting in the center of the hexagon. The powder x-ray diffraction performed on the crushed single crystals of Fe$_3$Ge,  as shown in Fig.~\ref{fig1}(b),  confirms the phase purity of the as-grown single crystals. The inset in Fig.~\ref{fig1}(b) depicts XRD pattern taken on the Fe$_3$Ge single crystal with the (0 0 $l$) Bragg's plane, suggesting that the crystal growth axis is parallel to the $c$-axis. In-plane ($\rho_{xx}$) and out-of-plane ($\rho_{zz}$) longitudinal electrical resistivity were measured as a function of temperature as shown in Figs.~\ref{fig1}(c) and ~\ref{fig1}(d), respectively,  confirming the metallic nature of Fe$_3$Ge throughout the measured temperature range. The electrical resistivity data on our Fe$_3$Ge are consistent with previous reports~\cite{Li2023,Zhang2024}. Further,  $\rho_{zz}$(T) was fitted using the Bloch-Wilson equation (see Eq.~\ref{Eq1}) for $n$=3, which accounts for the phonon contribution to the resistivity in combination with $T^2$ term to account for the electron-electron scattering~\cite{Minnigerode1983,Nandi2024}. On the other hand, for the best fitting of $\rho_{xx}$(T) data [see Fig.~\ref{fig1}(c)], we need to add an extra term $\rho_{sat}$ to Eq.~\ref{Eq1} (see Eq.~\ref{Eq2}) as $\rho_{xx}$(T) tends to saturate at higher temperatures~\cite{Wiesmann1977}. From these fittings, we obtained Debye temperature ($\Theta_D$) of about 460 K from $\rho_{xx}$ and 345 K from $\rho_{zz}$ data. The average Debye temperature of 402 K is close to the Debye temperature ($\Theta_D=$369 K) derived from the heat capacity measurements (see Fig. S2 in Ref.~\cite{Supple}).

\begin{equation}
\rho(T) = \rho_0 + A\left(\frac{T}{\Theta_D}\right)^3\int_{0}^{\frac{\Theta_D}{T}} \frac{x^3}{(e^x-1)(1-e^{-x})} \,dx + BT^2
\label{Eq1}
\end{equation}

\begin{equation}
\frac{1}{\rho_{tot}}= \frac{1}{\rho(T)}+\frac{1}{\rho_{sat}}
\label{Eq2}
\end{equation}

To elucidate the magnetic properties of Fe$_3$Ge, we measured magnetization as a function of temperature [$M(T)$] and field [$M(H)$] as shown in Fig.~\ref{fig2} for both $H\parallel z$ and $H\perp z$. From the $M(T)$ data [see Figs.~\ref{fig2}(a) and ~\ref{fig2}(b)], we find a spin-orientation transition at around $T_{SR}=$340 K below which the out-of-plane magnetization ($H\parallel z$) significantly reduces (35\% from the maximum at 340 K) ingoing to low-temperature below $T_{SR}$. On the other hand, the in-plane magnetization ($H\perp z$) does not decrease much (5\%) with decreasing temperature below $T_{SR}$. Further, we notice a ferromagnetic transition at a Curie temperature of $T_C=$680 K as demonstrated in the inset of Fig.~\ref{fig2}(b). The magnetisation isotherms $M(H)$ for $H\parallel z$ and $H\perp z$ at different sample temperatures are shown in Figs.~\ref{fig2}(c) and ~\ref{fig2}(d), respectively. The $M(H)$ data indicate that Fe$_3$Ge exhibits an in-plane ($H\perp z$) easy-axis of magnetization in its ground state, as evidenced by the inset shown in Fig.~\ref{fig2}(d). Interestingly, the $M(H)$ of all temperatures for $H\perp z$ do not change significantly as the saturation magnetization remains almost constant across all measured temperatures. In contrast, for $H\parallel z$, the $M(H)$ curves progressively shift to higher saturation magnetization values as the temperature decreases. This observation indicates that as the temperature decreases, the spins of Fe moments gradually reorient from the $z$-axis to the $xy$-plane, and eventually stabilize into an in-plane ferromagnetic configuration at low temperatures. This observation is consistent with the $M(T)$ data shown in Figs.~\ref{fig2}(a) and ~\ref{fig2}(b). Earlier also, it was reported that the Fe magnetic moments in this material exhibit a gradual deviation from the $z$-axis towards the $xy$-plane  below $T_{SR}$~\cite{Drijver1976,Albertini2004,Li2023}.

Next, the field-dependent Hall resistivity measured for in-plane ($\rho_{xy}$) and out-of-plane ($\rho_{zx}$) at various temperatures are presented in Figs.~\ref{fig2}(e) and ~\ref{fig2}(f), respectively. A sharp increase in the Hall resistivity at low fields, followed by saturation at higher fields, confirms the presence of anomalous Hall effect (AHE) in Fe$_3$Ge~\cite{Nagaosa2010,Zhang2024}. Figs.~\ref{fig2}(g) and ~\ref{fig2}(h) depict the Hall conductivity derived from the Hall resistivity  shown in Figs.~\ref{fig2}(e) and ~\ref{fig2}(f), respectively, using the relation,

 \begin{equation}
 \sigma_{xy/zx} = \frac{-\rho_{xy/zx}}{\rho_{xy/zx}^2 + \rho_{xx/zz}^2}
 \label{Eq3}
 \end{equation}
 where $\rho_{xx/zz}$ is the longitudinal resistivity measured along the $x/z$ direction of the crystal. \\

\begin{figure}[ht]
\includegraphics[width=1\linewidth, clip=true]{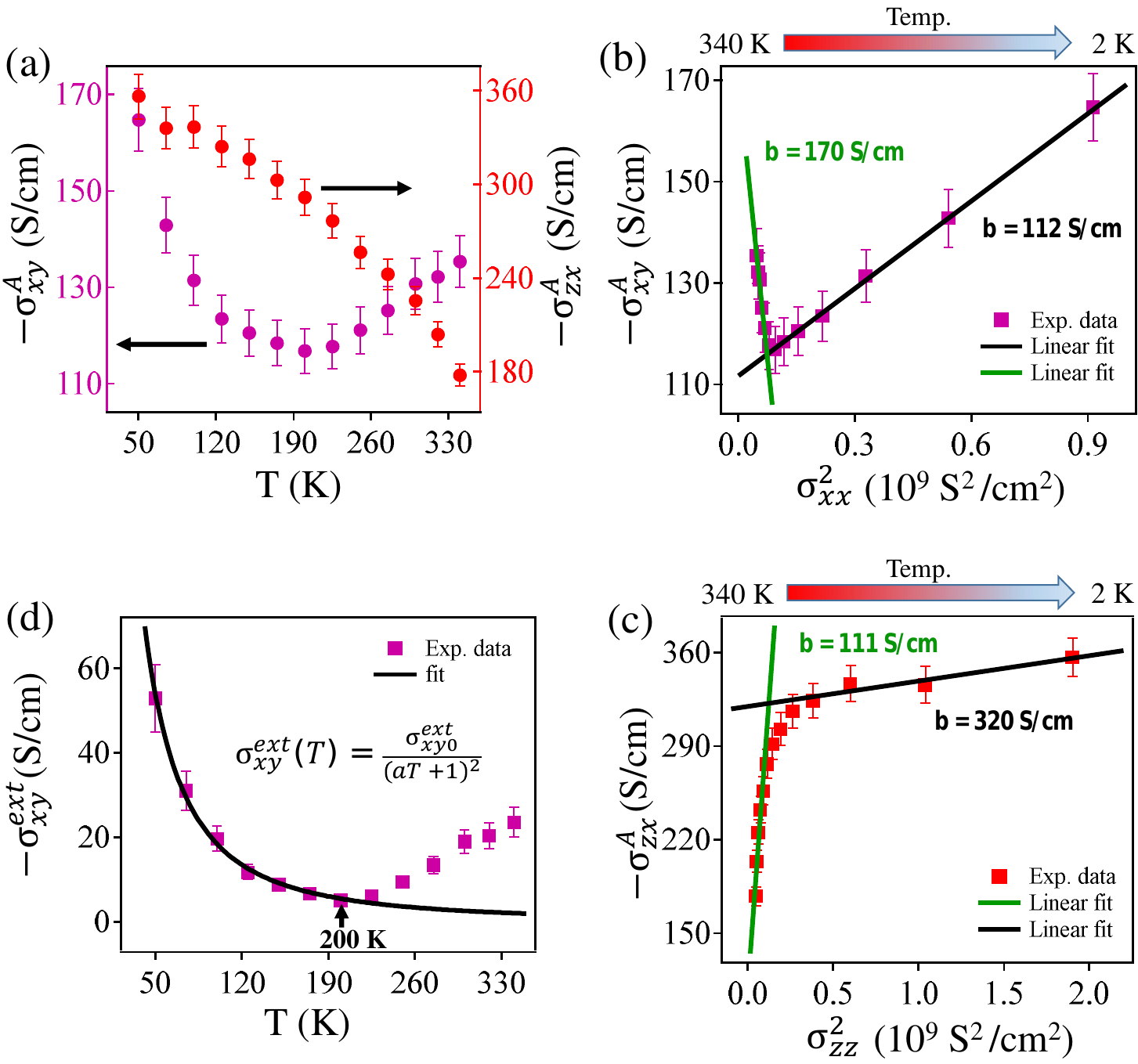}
\caption{(a) $\sigma_{xy}^A $ and $\sigma_{zx}^A $ plotted as a function of temperature. (b) Plot of \(-\sigma^{A}_{xy}\) $vs.$ \(\sigma^2_{xx}\). (c) Plot of \(-\sigma^{A}_{zx}\) $vs.$ \(\sigma^2_{zz}\). The solid lines in (b) and (c) are the linear fits by the Eq.~\ref{Eq5}. (d) Temperature-dependent in-plane extrinsic AHC (\(\sigma^{\text{ext}}_{xy}\)).}
\label{fig3}
\end{figure}

The total Hall resistivity of a ferromagnet can be written as $\rho_{H}=\rho_{H}^N+\rho_{H}^A = \mu_0R_0H + \mu_0R_SM$~\cite{Neubauer2009,Nagaosa2010},  where $R_0$ and $R_S$ are the normal and anomalous Hall coefficients [see Fig.~S3 in Ref.~\cite{Supple} for  $R_0(T)$ and $R_S(T)$ plots]. Thus, the field-dependent Hall resistivity of $\rho_{xy}(H)$ and $\rho_{zx}(H)$ measured at various temperatures, as shown in Figs.~\ref{fig2}(e) and ~\ref{fig2}(f),   were fitted using the total resistivity equation and extracted the anomalous Hall resistivity $\rho^A_{xy}$ and $\rho^A_{zx}$ [see Fig.~S3 in Ref.~\cite{Supple}]. Finally,  from the fits we converted $\rho_{xy}(H)$ and $\rho_{zx}(H)$ into the anomalous Hall conductivity $\sigma^A_{xy}$ and $\sigma^A_{zx}$ and plotted as a function temperature as shown in Fig.~\ref{fig3}(a).  A unified scaling relation between the anomalous Hall resistivity ($\rho^A_{H}$) and the longitudinal electrical resistivity ($\rho$)  remains elusive. Different functional forms have been reported for $\rho^A_H$ as a function of $\rho$, $\rho^A_H=f(\rho)$. Initially, Smit $\it{et.~al.}$ attributed the anomalous Hall effect to the extrinsic skew-scattering of electrons with magnetic impurities and proposed that $\rho_{sk}^A\propto \rho_{xx}$~\cite{smit1955}. On the other hand, Berger $\it{et.~al.}$ suggested that the extrinsic side-jump scattering of electrons with magnetic impurities could also contribute to the AHE such as $\rho_{sj}^A\propto \rho_{xx}^2$~\cite{berger1970}. In contrast, Karplus and Luttinger (KL) proposed an intrinsic contribution to the anomalous Hall effect, arising from the strong spin-orbit interactions combined with inter-band scattering, leading to  $\rho_{int}^A\propto \rho_{xx}^2$~\cite{Karplus1954, Nagaosa2010}. The KL mechanism was recently revisited through the Berry phase mechanism, and it realized that the intrinsic Hall effect is a dominant contributor to the anomalous Hall effect~\cite{Nagaosa2010}. In this regard, the anomalous Hall conductivity can be expressed as some of all three contributions following Matthiessen's rule,

  \begin{figure}[ht]
	\includegraphics[width=\linewidth, clip=true]{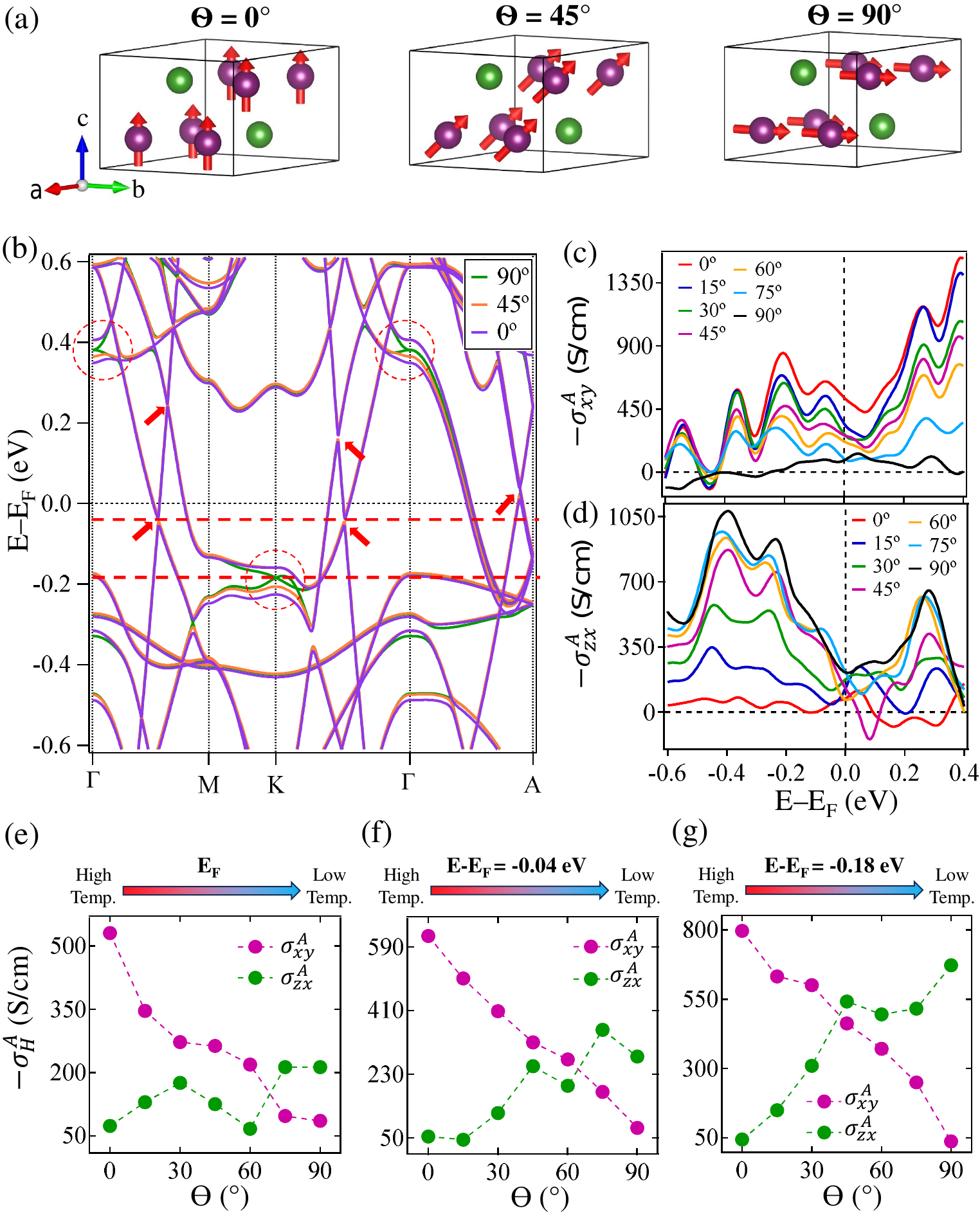}
	\caption{(a) Schematic representation of various spin configurations in $Fe_3Ge$. (b) Electronic band structure calculated using DFT for various spin configurations as depicted in (a)  including the spin-orbit coupling effects. (c) In-plane ($\sigma_{xy}^A$) and (d) out-of-plane ($\sigma_{zx}^A$) intrinsic anomalous Hall conductivity,  plotted as a function of binding energy, calculated for various spin configurations. (e)-(g) Intrinsic AHC plotted as a function of angle ($\Theta$) at $E_F$, $E-E_F=-0.04$ eV (Weyl-node), and $E-E_F=-0.18$ eV (Dirac-node).}
	\label{fig4}
\end{figure}

\begin{equation}
 \sigma_{AH} = \sigma_{int} + \sigma_{sk} + \sigma_{sj},
 \label{Eq4}
\end{equation}

where $\sigma_{int}$ is the intrinsic Hall contribution,  $\sigma_{sk}$ is the extrinsic skew-scattering contribution, and $\sigma_{sj}$ is the extrinsic side-jump contribution. \\

Usually, intrinsic and extrinsic Hall contributions coexist, and distinguishing the intrinsic from the extrinsic contribution is a major experimental challenge. In this regard, Tian $et.~ al.$~\citep{Tian2009} proposed a novel scaling law (TYJ) for the anomalous Hall effect, which offered a more reliable and cohesive framework for comprehending the relation among various AHE contributions. The TYJ scaling law incorporates residual resistivity ($\rho_{xx0}$) as $\rho_{xy}^A=f(\rho_{xx0},\rho_{xx})$. Thus, the empirical form of the TYJ model for AHE is given by,

\begin{equation}
 -\sigma_{xy}^A(T)=\rho_{xy0}^{ext}\sigma_{xx}^2(T)+b(T),
 \label{Eq5}
 \end{equation}

where $\sigma_{xx0}=1/\rho_{xx0}$ is the residual conductivity and $b(T)$ is temperature-dependent intrinsic Hall conductivity arising from the Berry curvature.

Fig.~\ref{fig3}(b) depicts the plot of $\sigma_{xy}^A$ $vs.$ $\sigma_{xx}^2$, overlapped with fits (solid lines) using Eq.~\ref{Eq5}. From Fig.~\ref{fig3}(b), it is evident that $\sigma_{xy}^A$ shows differing linear dependencies on  $\sigma_{xx}^2$ in different temperature regions. As a result, we get the intrinsic Hall conductivity of $b=170$ S/cm in the high-temperature region, which is 112 S/cm in the low-temperature region. Similarly, from the plot of $\sigma_{zx}^A$ $vs.$ $\sigma_{zz}^2$ [see Fig.~\ref{fig3}(c)] we observe $b=111$ S/cm  in the high-temperature region and 320 S/cm in the low-temperature region. These observations demonstrate the temperature dependence of intrinsic Hall contribution to the total Hall conductivity. This conclusion is further supported by the plot of $\rho_{xy/zx}^A/M_s$ $vs.$ $\rho^2_{xx/zz}$ as depicted in Ref.~\cite{Supple} [see Fig. S3(f)], a deviation in the linear fit confirms the temperature-dependent intrinsic AHC, otherwise perfect linear behaviour is anticipated between $\rho^A$ and $\rho^2$ for the temperature independent intrinsic AHC~\cite{Tian2009, Singh2024}. Importantly, in the high-temperature region, the in-plane intrinsic contribution dominates the out-of-plane. In contrast, the out-of-plane intrinsic contribution dominates the in-plane in the low-temperature region.

Next, the extrinsic Hall conductivity can be isolated by subtracting the intrinsic contribution from the total Hall conductivity. Since the intrinsic Hall conductivity for $\sigma_{xy}$ does not change much within the measured temperature range, the extrinsic contribution ($\sigma_{xy}^{ext}$) is obtained by subtracting the low-temperature intrinsic AHC of $b=$112 S/cm. The resultant $\sigma_{xy}^{ext}$ is plotted as a function of temperature in Fig.~\ref{fig3}(d), from which one can notice that with increasing temperature, its magnitude decreases significantly. Therefore, the sharp increase in total anomalous Hall conductivity below 200 K can be attributed to the extrinsic contribution. Note that the increase in AHE above 200 K is due to the high-temperature intrinsic AHC. In ferromagnetic metals, the value of $\frac{\epsilon_{SO}}{E_F}$ is of the order of $\sim 10^{-2}$ and  $\frac{e^2}{ha'}\approx7.97\times10^2~S/cm$ for an average lattice constant of the studied system, $a'=(2a+c)/3=4.86 ~\AA$~\cite{Wang2016, Roy2020}. Thus, the extrinsic anomalous Hall conductivity due to the side-jump is given by $\sigma_{sj}\approxeq\frac{e^2}{ha'}(\frac{\epsilon_{SO}}{E_F})\approx7.97~S/cm$ which is negligibly small. Therefore, the extrinsic skew-scattering mechanism dominates at low temperatures where the inelastic electron-electron scattering is minimal, and the sample's longitudinal conductivity is high [the clean limit regime]~\cite{Onoda2006}. However, as the temperature increases, the probability of inelastic scattering increases because of the electron-phonon interactions, reducing the extrinsic skew-scattering contribution to the AHC~\cite{Low2023}.

Shitade $\it{et. ~al.}$ had introduced a model that elucidates the impact of electron-phonon interactions on the skew-scattering contribution of the anomalous Hall conductivity~\cite{Shitade2012}. Since in our studied system (Fe$_3$Ge) also we observe a strong electron-phonon interactions within the temperature range of 50 and 200 K [see Fig.~\ref{fig1}(c)], using the inelastic scattering rate ($\gamma$) as implemented in Ref.~\cite{Shitade2012}, we could fit the extrinsic skew-scattering Hall conductivity by the equation $\sigma_{xy}^{ext}=\frac{\sigma_{xy0}^{ext}}{(\gamma/\gamma_0+1)^2}=\frac{\sigma_{xy0}^{ext}}{(aT+1)^2}$ as shown in Fig.~\ref{fig3}(d). Here, $\gamma/\gamma_0=aT$ for the electron-phonon scattering, and $a$ is a constant. The same procedure cannot be applied for $\sigma_{zx}^A$ data to extract the extrinsic Hall contribution as the intrinsic AHC values change significantly from the high-temperatures (111 S/cm) to the low-temperatures (320 S/cm).

To further understand the anomalous Hall conductivity in Fe$_3$Ge, we performed $ab-initio$ first-principles calculations as shown in Fig.~\ref{fig4}. Since, experimentally, we notice that the magnetic easy-axis rotates from out-of-plane towards in-plane in-going from high-temperature to low-temperature, we performed the calculations by fixing the magnetic easy-axis at different angles ($\Theta$) with respect to the crystallographic $c$-axis. $\Theta=0^o$ means the magnetic easy-axis is parallel to the $c$-axis and 90$^o$ means the magnetic easy-axis is perpendicular to the $c$-axis as demonstrated in the Fig.~\ref{fig4}(a). Fig.~\ref{fig4}(b) depicts the electronic band structure of Fe$_3$Ge calculated including the spin-orbit coupling effects for the magnetic configurations of $\Theta=0^o$, $45^o$, and $90^o$. From Fig.~\ref{fig4}(b), we can identify several Weyl points near the Fermi level that are gapped out due to spin-orbit coupling. In addition, we observe a couple of Dirac-like band crossings above and below the Fermi level at $\Gamma$ and $K$ high-symmetry points, respectively. Interestingly, the degeneracy of the Dirac point is intact for the  $\Theta=90^o$ magnetic configuration, i.e., the easy-magnetic axis is perpendicular to the $c$-axis. However, moving the easy-axis towards the $c$-axis, the degeneracy of the Dirac points is lifted by opening a gap at the nodes. A maximum Dirac nodal gap is observed for the $\Theta=0^o$ magnetic configuration.

Figs.~\ref{fig4}(c) and  ~\ref{fig4}(d) show the calculated anomalous Hall conductivity of in-plane ($\sigma^A_{xy}$) and out-of-plane ($\sigma^A_{zx}$), respectively,  plotted as a function of binding energy for various spin configurations. $\sigma_{xy}^A$ and $\sigma_{zx}^A$ were calculated with the magnetization vector kept along the [0001] ($z$) and [01$\bar{1}$0]($y$) directions, respectively. Figs.~\ref{fig4}(e)-\ref{fig4}(g) show $\sigma^A_{xy}$ and $\sigma^A_{zx}$ plotted as a function spin-configuration angle at the Fermi level ($E_F$), the Weyl point ($E_F$-0.04 eV),  and  Dirac point ($E_F$-0.18 eV), respectively. From Figs.~\ref{fig4}(e)-\ref{fig4}(g), it is evident that the in-plane AHC ($\sigma_{xy}^A$) dominates the out-of-plane AHC ($\sigma_{zx}^A$) above $T_{SR}$ ($\Theta=0^o$). However, as the temperature decreases ($\Theta$ increases), the $\sigma_{xy}^A$ ($\sigma_{zx}^A$) gradually decreases (increases). Though this behavior is similar at $E_F$, Weyl, and Dirac-point energy positions, $\sigma_{zx}^A$ dominance over $\sigma_{xy}^A$ at low temperatures is significantly higher at the Weyl and Dirac-points. Further, the dominance cross-over between  $\sigma_{xy}^A$  and  $\sigma_{zx}^A$ ingoing from $T_{SR}$ to low-temperature is qualitatively in agreement with the experimental observations as demonstrated in Figs.~\ref{fig3}(b) and ~\ref{fig3}(c).

Thus, in Fe$_3$Ge, the direction of magnetization has a significant impact on the electronic band structure, providing a means to tune the electronic properties by controlling the magnetization orientation. Specifically, the nature of Dirac and Weyl points is highly sensitive to the magnetization angle. As illustrated in Figs.~\ref{fig4}(b) and ~\ref{fig4}(c), the band gap at the Dirac point located at the high-symmetry $K$ point exhibits significant variation with changes in the magnetization direction. The direction of magnetization is determined by the easy-magnetic axis, which gradually changes with temperature, as evidenced by the experimental data [see Figs.~\ref{fig2}(a) and ~\ref{fig2}(b)]. As the easy-axis reorients, the magnetization direction evolves, leading to modifications in the Berry curvature. This, in turn, results in a temperature-dependent intrinsic AHC.

\section{Conclusions}
In conclusion, we conducted a comprehensive investigation on Fe$_3$Ge for its anomalous Hall effect behavior along the in-plane and out-of-plane directions. Our study reveals significant anisotropy in the anomalous Hall effect (AHE). Notably, the gradual spin reorientation from the out-of-plane to in-plane direction, from high-temperature to low-temperature in Fe$_3$Ge, provides a unique opportunity to study the temperature dependence on the intrinsic Hall conductivity experimentally. The theoretical predictions qualitatively support our experimental results.

\section{Acknowledgement}\label{5}

S.G. acknowledges the University Grants Commission (UGC), India, for the Ph.D. fellowship. S.T. thanks the Science Engineering Research Board (SERB), Anusandhan National Research Foundation (ANRF), India, for the financial support through Grant no. CRG/2023/00748. This research has used the Technical Research Centre (TRC) Instrument facilities of S. N. Bose National Centre for Basic Sciences, established under the TRC project of the Department of Science and Technology (DST), Govt. of India.

\section{Data Availability}
 The data supporting this study’s findings are available  within the article.

\bibliography{Fe3Ge}

\section{Supplemental Information}

\begin{figure}[!ht]
	\includegraphics[width=\linewidth, clip=true]{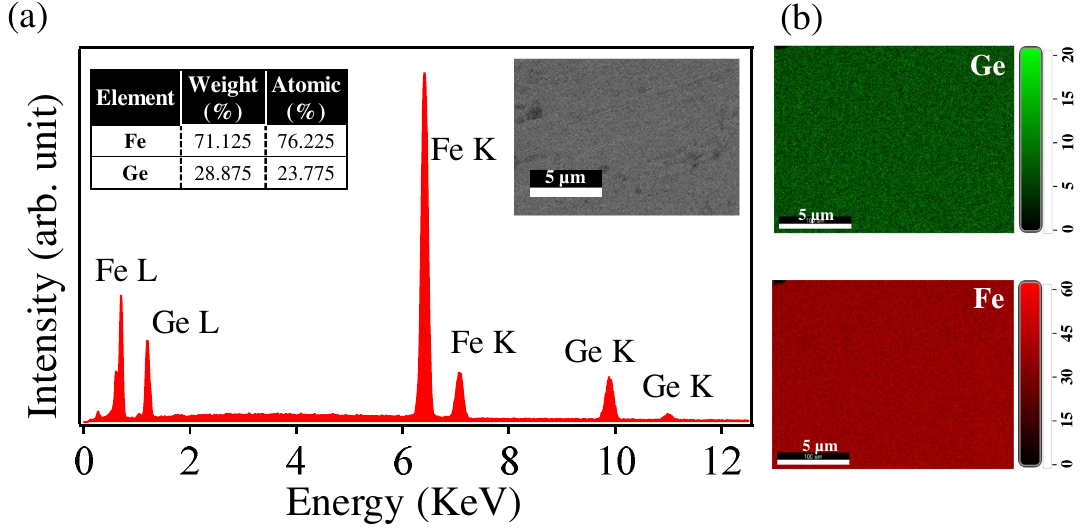}
	\caption{Fig. S1: (a) EDS spectra of Fe\(_3\)Ge and inset shows the tabulated elemental ratios present in the system. (b) The elemental mapping of Fe and Ge in the studied single crystals.}
	\label{figS1}
\end{figure}

\begin{figure*}[!ht]
	\includegraphics[width=0.7\linewidth, clip=true]{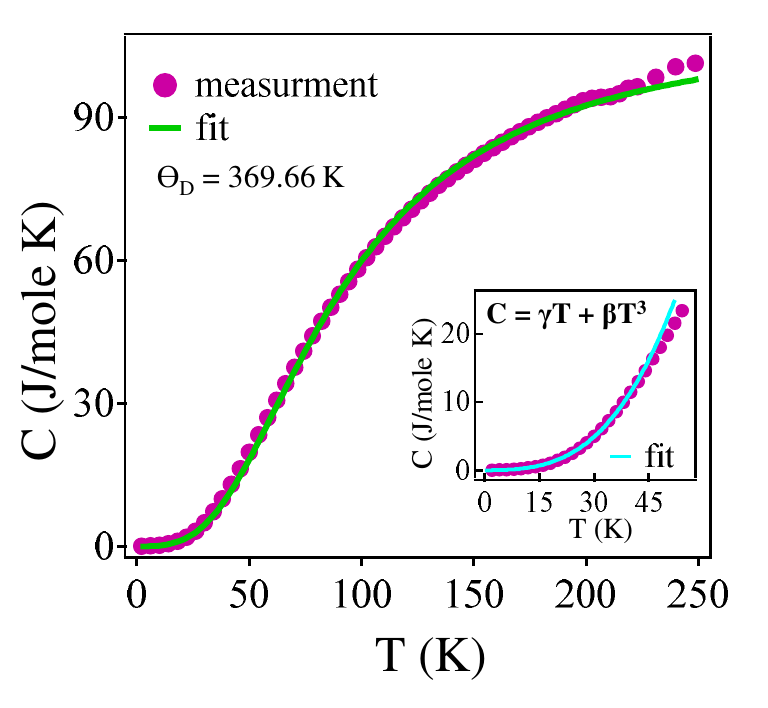}
	\caption{Fig. S2: Temperature-dependent specific heat. Solid line is a fit with Debye model as discussed in the . The inset shows a zoomed-in image of the low-temperature region.}
	\label{figS2}
\end{figure*}

  \begin{figure*}[hb]
	\includegraphics[width=0.9\linewidth, clip=true]{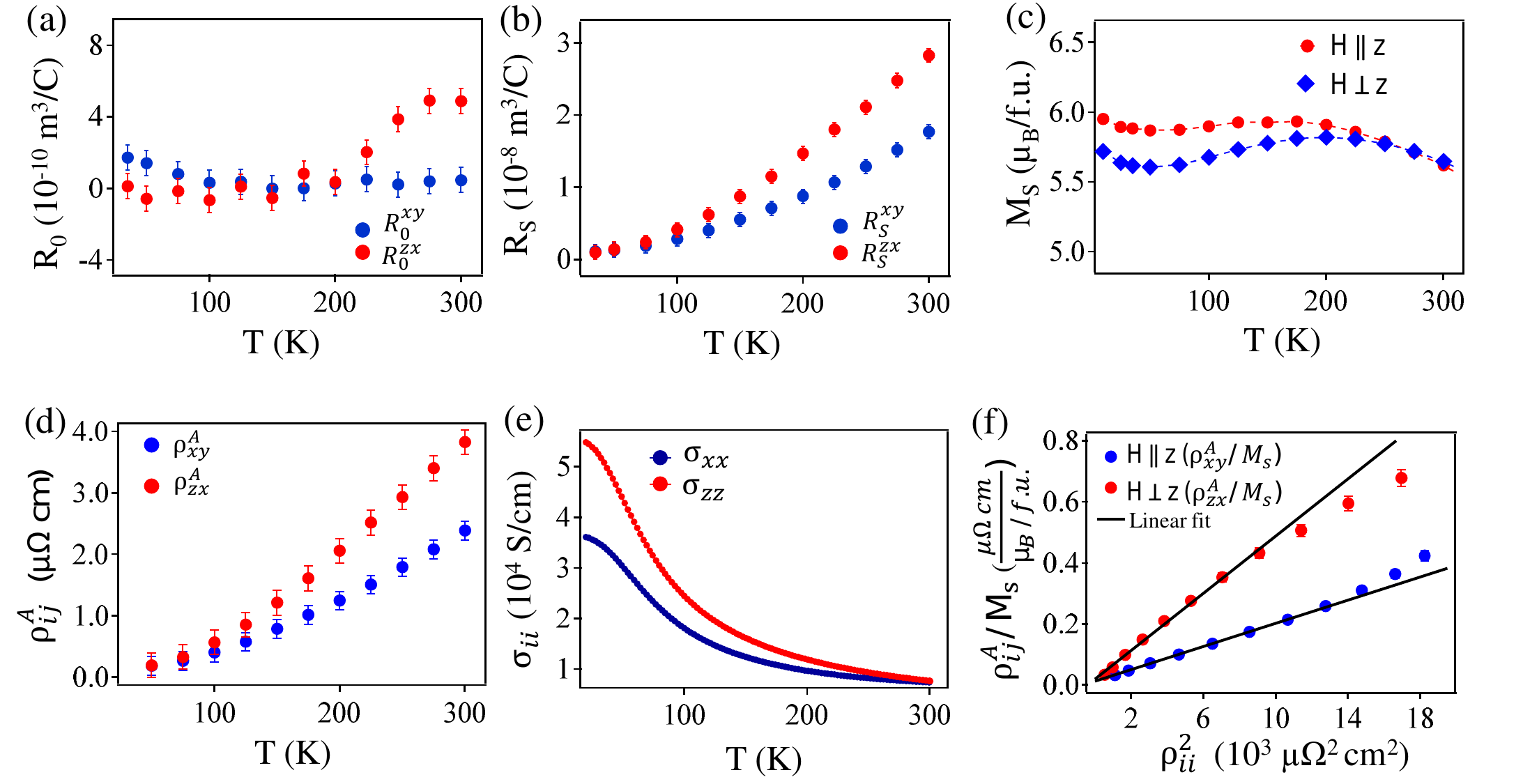}
	\caption{Fig. S3: Panels (a) and (b) depict the temperature evolution of the normal Hall coefficient (\(R_0\)) and anomalous Hall coefficient (\(R_S\)), respectively.  (c) Saturation magnetization ($M_S$) is  plotted as functions of temperature.  (d) Anomalous Hall resistivity and (e) longitudinal conductivity are plotted as a function of temperature. (f) Plot of \(\rho^{A}_{zx/xy}/M_s\) $vs.$ \(\rho^2_{zz/xx}\), with the solid line depicting a linear fit.}
	\label{figS3}
\end{figure*}

  \begin{figure*}[hb]
	\includegraphics[width=0.9\linewidth, clip=true]{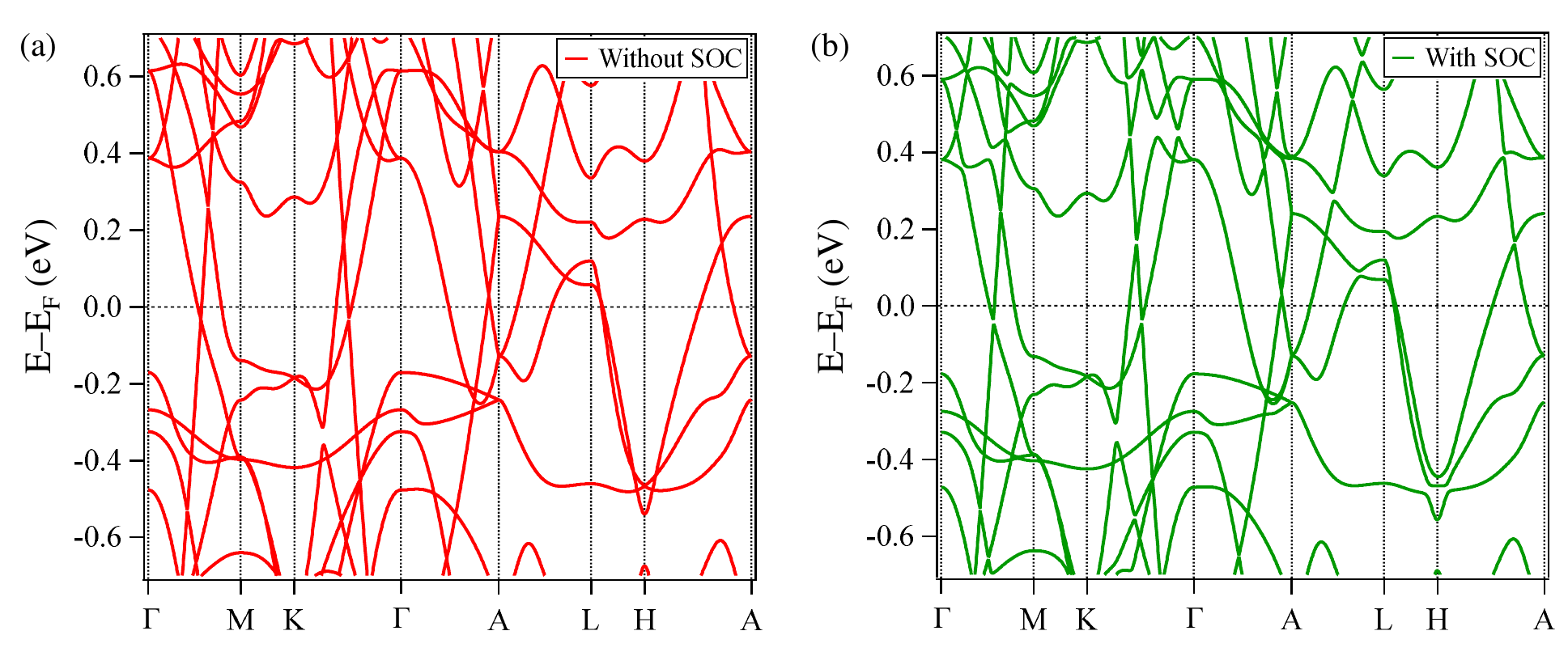}
	\caption{Fig. S4: Electronic band structure of Fe$_3$Ge calculated (a) without including SOC and (b) with SOC while taking the easy-magnetic axis perpendicular to the crystallographic $c$-axis.}
	\label{figS4}
\end{figure*}

In Fig.~\ref{figS2}(c),  specific heat plotted as a function of temperature for Fe$_3$Ge. The Debye model is used to fit the specific heat data using the Eq.~\ref{EqS1}. From the fitting we obtained a temperature of $\Theta_D=$369 K.

\begin{equation}
C (T) = 9nR\left(\frac{T}{\Theta_D}\right)^3\int_{0}^{\frac{\Theta_D}{T}} \frac{e^xx^4}{(e^x-1)^2} \,dx.
\label{EqS1}
\end{equation}

Additionally, from the fitting we obtained $9nR = 311.80$ $ J$ $mole^{-1}$ $K^{-1}$ which results in $3nR = 103.93$. These values are in good agreement with the projected Dulong-Petit limit of $3nR = 99.72$ J $mol^{-1}$ $K^{-1}$, where $n$ is the number of atoms per formula unit in the compound ($n$ = 4 for Fe$_3$Ge) and $R = 8.31 $ J $mol^{-1}$ $K^{-1}$. Further, specific heat at low temperatures can be expressed best by using the Eq.~\ref{EqS2}.

\begin{equation}
C = \gamma T + \beta T^3
\label{EqS2}
\end{equation}

Here,  $\gamma T$ represents the electronic contribution and $\beta T^3$ the phononic contribution to the specific heat. The obtained $\gamma$ and $\beta$ values are 2.98  mJ $mol^{-1}$ $K^{-2}$  and 0.18 mJ $mol^{-1}$ $K^{-4}$, respectively.

\end{document}